\documentclass[aps,prl, floatfix, tightenlines ,superscriptaddress,showpacs,preprintnumbers,amsmath,amssymb, reprint,letter]{revtex4-1}
\usepackage{graphicx} 
\usepackage{dcolumn}  

\usepackage{amstext}
\usepackage{color}
\graphicspath{{ps}}

\usepackage{verbatim}

\begin{document}

\newcommand{\tauToKSpi} {\tau^\pm  \rightarrow     K^0_S        \pi^\pm      \nu_\tau}
\newcommand{\tauToKSpiM}{\tau^-     \rightarrow     K^0_S        \pi^-    \nu_\tau}
\newcommand{\tauToKSpiP}{\tau^+     \rightarrow     K^0_S        \pi^+    \bar{\nu}_\tau}

\newcommand{\ACPth}{A^\text{cp}}
\newcommand{\ACPthi}{A^\text{cp}_{i} }
\newcommand{\ACPsig}{A^\text{cp}}
\newcommand{\ACPsigi}{A^\text{cp}_{i} }

\newcommand{\ReETA}{\Re(\eta_S)}
\newcommand{\ImETA}{\Im(\eta_S)}
\newcommand{\IMcs}{\Im}


\preprint{\vbox{ \hbox{   }
                         \hbox{BELLE Preprint 2010-27}
                         \hbox{Phys. Rev. Lett. 107, 131801 (2011)}
}}

\title{Search for CP violation in $\tau^\pm  \rightarrow K^0_S \pi^\pm \nu_\tau$ decays at Belle}

\affiliation{Budker Institute of Nuclear Physics, Novosibirsk}
\affiliation{Faculty of Mathematics and Physics, Charles University, Prague}
\affiliation{Chiba University, Chiba}
\affiliation{University of Cincinnati, Cincinnati, Ohio 45221}
\affiliation{Justus-Liebig-Universit\"at Gie\ss{}en, Gie\ss{}en}
\affiliation{The Graduate University for Advanced Studies, Hayama}
\affiliation{University of Hawaii, Honolulu, Hawaii 96822}
\affiliation{Hiroshima Institute of Technology, Hiroshima}
\affiliation{University of Illinois at Urbana-Champaign, Urbana, Illinois 61801}
\affiliation{Institute of High Energy Physics, Chinese Academy of Sciences, Beijing}
\affiliation{Institute of High Energy Physics, Vienna}
\affiliation{Institute of High Energy Physics, Protvino}
\affiliation{Institute for Theoretical and Experimental Physics, Moscow}
\affiliation{J. Stefan Institute, Ljubljana}
\affiliation{Kanagawa University, Yokohama}
\affiliation{Institut f\"ur Experimentelle Kernphysik, Karlsruher Institut f\"ur Technologie, Karlsruhe}
\affiliation{Korea Institute of Science and Technology Information, Daejeon}
\affiliation{Korea University, Seoul}
\affiliation{Kyungpook National University, Taegu}
\affiliation{\'Ecole Polytechnique F\'ed\'erale de Lausanne (EPFL), Lausanne}
\affiliation{Faculty of Mathematics and Physics, University of Ljubljana, Ljubljana}
\affiliation{University of Maribor, Maribor}
\affiliation{Max-Planck-Institut f\"ur Physik, M\"unchen}
\affiliation{University of Melbourne, School of Physics, Victoria 3010}
\affiliation{Nagoya University, Nagoya}
\affiliation{Nara Women's University, Nara}
\affiliation{National United University, Miao Li}
\affiliation{Department of Physics, National Taiwan University, Taipei}
\affiliation{H. Niewodniczanski Institute of Nuclear Physics, Krakow}
\affiliation{Nippon Dental University, Niigata}
\affiliation{Niigata University, Niigata}
\affiliation{University of Nova Gorica, Nova Gorica}
\affiliation{Novosibirsk State University, Novosibirsk}
\affiliation{Osaka City University, Osaka}
\affiliation{Saga University, Saga}
\affiliation{University of Science and Technology of China, Hefei}
\affiliation{Seoul National University, Seoul}
\affiliation{Sungkyunkwan University, Suwon}
\affiliation{School of Physics, University of Sydney, NSW 2006}
\affiliation{Tata Institute of Fundamental Research, Mumbai}
\affiliation{Excellence Cluster Universe, Technische Universit\"at M\"unchen, Garching}
\affiliation{Tohoku Gakuin University, Tagajo}
\affiliation{Tohoku University, Sendai}
\affiliation{Department of Physics, University of Tokyo, Tokyo}
\affiliation{Tokyo Metropolitan University, Tokyo}
\affiliation{Tokyo University of Agriculture and Technology, Tokyo}
\affiliation{CNP, Virginia Polytechnic Institute and State University, Blacksburg, Virginia 24061}
\affiliation{Wayne State University, Detroit, Michigan 48202}
\affiliation{Yonsei University, Seoul}
\author{M.~Bischofberger}\affiliation{Nara Women's University, Nara} 
\author{H.~Hayashii}\affiliation{Nara Women's University, Nara} 
  \author{K.~Adamczyk}\affiliation{H. Niewodniczanski Institute of Nuclear Physics, Krakow} 
  \author{H.~Aihara}\affiliation{Department of Physics, University of Tokyo, Tokyo} 
  \author{V.~Aulchenko}\affiliation{Budker Institute of Nuclear Physics, Novosibirsk}\affiliation{Novosibirsk State University, Novosibirsk} 
 \author{A.~M.~Bakich}\affiliation{School of Physics, University of Sydney, NSW 2006} 
  \author{V.~Balagura}\affiliation{Institute for Theoretical and Experimental Physics, Moscow} 
  \author{E.~Barberio}\affiliation{University of Melbourne, School of Physics, Victoria 3010} 
  \author{K.~Belous}\affiliation{Institute of High Energy Physics, Protvino} 
  \author{A.~Bozek}\affiliation{H. Niewodniczanski Institute of Nuclear Physics, Krakow} 
% ...
\collaboration{The Belle Collaboration}
\noaffiliation

\begin{abstract}
We report on a search for CP violation in $\tau^\pm \rightarrow K^0_S \pi^\pm \nu_\tau$ decays using a data sample of $699\,\mathrm{fb}^{-1}$ collected by the Belle experiment at the KEKB electron-positron asymmetric-energy collider.  The CP asymmetry is measured in four bins of the invariant mass of the $K^0_S\pi^\pm$ system and found to be compatible with zero with a precision of O($10^{-3}$) in each mass bin. Limits for the CP violation parameter $\Im(\eta_S)$ are given at the $90\%$ confidence level. These limits are $|\Im(\eta_S)|<0.026$ or better, depending on the parameterization used to describe the hadronic form factors, and improve upon previous limits by one order of magnitude. 
\end{abstract}

\maketitle


{\renewcommand{\thefootnote}{\fnsymbol{footnote}}}
\setcounter{footnote}{0}

To date CP violation (CPV) has been observed only in the $K$ and $B$ meson systems. In the Standard Model (SM), all observed CPV effects can be explained by the irreducible complex phase in the Cabibbo-Kobayashi-Maskawa (CKM) quark mixing matrix \cite{Kobayashi:1973fv}.
To find new physics, it is important to look for other CP-violating effects in as many systems as possible. One such system is the $\tau$ lepton. In hadronic $\tau$ decays, one can search for CPV effects of possible new physics that could originate, for example, from the Minimal Supersymmetric Standard Model \cite{Christova:2002sw, Ibrahim:2007fb} or from multi-Higgs-doublet models \cite{Weinberg:1976hu,Grossman:1994jb} that play an important role in strangeness changing processes. 

This paper describes a search for CPV in $\tauToKSpi$ decays. It should be noted that CPV in $K^0$ decays leads to a small SM CP asymmetry of O($10^{-3}$) in the rates of this $\tau$ decay mode \cite{Bigi:2005ts,Calderon:2007rg}. This asymmetry is just below our experimental sensitivity.
Here the focus will be on CPV that could arise from a charged scalar boson exchange \cite{Kuhn:1996dv}, e.g., a charged Higgs boson. This type of CPV cannot be observed from measurement of $\tau^\pm$ decay rates. However, it can be detected as a difference in the $\tau^\pm$ decay angular distributions and is accessible without requiring information about the $\tau$ polarization or the determination of the $\tau$ rest frame.
Limits for the CPV parameter in this decay mode have been published previously by the CLEO collaboration from an analysis of $13.3\,\mathrm{fb}^{-1}$ of data \cite{Bonvicini:2001xz}.

In the SM, the differential decay width in the hadronic rest frame ($\vec{q}_1+\vec{q}_2=0$) is given by (see \cite{Kuhn:1996dv} for details)
\begin{align}
d\Gamma_{\tau^-} &=  \frac{G_F^2}{2m_{\tau}}\sin^2\theta_c \frac{1}{(4\pi)^3} \frac{(m_\tau^2-Q^2)^2}{m_\tau^2}|\vec{q}_1| \label{eqn:DecayWidth}\\
&\times\frac{1}{2}\left( \displaystyle \sum_{X} \bar{L}_X W_X  \right)\frac{dQ^2}{\sqrt{Q^2}}\frac{d\!\cos\theta}{2}\frac{d\alpha}{2\pi}\frac{d\!\cos\beta}{2},\nonumber
\end{align}
where $G_F$ is the Fermi coupling constant, $\theta_c$ is the Cabibbo angle, $m_\tau$ is the mass of the $\tau$ lepton,  $\vec{q}_1$ and $\vec{q}_2$ denote the three-momenta of the $K^0_S$ and $\pi^-$, respectively, and $Q^2 = (q_1 + q_2)^2$ is the square of the invariant mass of the $K^0_S\pi^\pm$ system.
The four hadronic functions $W_X$ with $X \in (B,SA,SF,SG)$ (see \cite{Kuhn:1992nz}) are formed from the vector and scalar form factors $F(Q^2)$ and $F_S(Q^2)$ and are proportional to $|F|^2$,  $|F_S|^2$, $\Re(FF_S)$, and $\Im(FF_S)$, respectively. The $L_X$ functions, which contain the angular dependence, can be calculated from electroweak theory (see \cite{Kuhn:1996dv}).
The angle $\beta$ is defined by  $\cos\beta = \vec{n}_L\cdot \hat{q}_1$ where $\hat{q}_1 = \vec{q}_1/|\vec{q}_1|$ is the direction of the $K^0_S$ and $\vec{n}_L$ is the direction of the $e^+e^-$ center of mass (CM) system, both observed in the hadronic rest frame. The azimuthal angle $\alpha$ is not observable in this experiment and has to be integrated over. 
The variable $\theta$ is the angle between the direction opposite to the direction of the CM system and the direction of the hadronic system in the $\tau$ rest frame.
In this experiment, the direction of the $\tau$ is not known but $\theta$ can be calculated from the hadronic energy $E_h$ measured in the CM system:
\begin{equation}
\cos\theta = \frac{2xm_\tau^2-m_\tau^2-Q^2}{(m_\tau^2-Q^2)\sqrt{1-4m_\tau^2/s}}, \quad x=2\frac{E_h}{\sqrt{s}},
\label{eqn:costheta}
\end{equation}
where $s=4E_{\mathrm{beam}}^2$ denotes the squared CM energy.

The effect of the exchange of a charged scalar boson can be introduced by replacing the scalar form factor $F_S$ with 
\begin{equation}
F_S(Q^2) \rightarrow \tilde{F_S}(Q^2) = F_S(Q^2)+\frac{\eta_S}{m_{\tau}}F_H(Q^2), \label{Eq:FStilde}
\end{equation}
where $F_H$ denotes the form factor for the scalar boson exchange [$F_H = \langle K^0(q_1)\pi^-(q_2)| \bar{u}s | 0 \rangle$] and $\eta_S$ is the corresponding dimensionless complex coupling constant \cite{Kuhn:1996dv, Kiers:2008mv, Choi:1994ch}.
The differential decay width for the CP conjugate process,  $d\Gamma_{\tau^+}$, is obtained from Eq.~(\ref{eqn:DecayWidth}) and Eq.~(\ref{Eq:FStilde}) by the replacement $\eta_S \rightarrow \eta_S^*$. Using this relation the CP violating quantity is given by \cite{Kuhn:1996dv}
\begin{align}
\displaystyle \Delta_{LW} & \equiv  \frac{1}{2} \left[ \sum_X  \bar{L}_X W_X(\eta_S)  -  \sum_X\bar{L}_X W_X(\eta_S^*) \right]  \label{eqn:deltaLW}\\
& =   -4  \frac{m_\tau}{\sqrt{Q^2}}|\vec{q}_1| {\IMcs(FF_H^*) \ImETA\cos{\psi}\cos{\beta} }, \nonumber
\end{align}
where  $\psi$ denotes the angle between the direction of the CM frame and the direction of the $\tau$ as seen from the hadronic rest frame and can be calculated as
\begin{equation}
\cos\psi = \frac{x(m_\tau^2+Q^2)-2Q^2}{(m_\tau^2-Q^2)\sqrt{x^2-4Q^2/s}}.
\label{eqn:cospsi}	
\end{equation}
Since the CP violating term is proportional to $\cos\beta\cos\psi$, it cancels out if one integrates over the angles $\beta$ and $\psi$, e.g., for branching fractions. Furthermore, the CP violating effect is only observable if $\IMcs(FF_H^*)\neq 0$.  The form factor $F_H$ is related to the SM weak scalar form factor $F_S$ via:
\begin{equation} 
F_H(Q^2) = \frac{Q^2}{m_u-m_s}F_S(Q^2) \label{eqn:FHfromFS}
\end{equation}
where $m_u$ and $m_s$ denote the up and strange quark masses, respectively. The derivation of Eq.~(\ref{eqn:FHfromFS}) is discussed in \cite{Kuhn:1996dv} although $F_H$ is not used there explicitly. The chosen value $(m_u-m_s)=-0.1\,\mathrm{GeV}/c^2$ defines the scale of the CPV parameter $\Im(\eta_S)$. Because the CLEO collaboration used a different relation $F_H =M F_S$ with $M=1\,\mathrm{GeV}/c^2$ as well as a different normalization of $F_S(Q^2)$, $\Im(\eta_S)$ is not the same as the CP parameter $\Lambda$ that was used in \cite{Bonvicini:2001xz}. In the following, the approximate relation $\ImETA \simeq -1.1 \Lambda$ is used to enable a comparison of the results.
 
To extract the CP violating term in Eq.~(\ref{eqn:deltaLW}), we define an asymmetry in bin $i$ of $Q^2$ using the difference of the differential $\tau^+$ and $\tau^-$ decay widths weighted by $\cos\beta\cos\psi$:  
\begin{align}
\displaystyle \ACPthi &= \frac{\iiint_{Q^2_{1,i}}^{Q^2_{2,i}}   \cos{\beta} \cos{\psi}  \left( \frac{d\Gamma_{\tau^-}}{d\omega} - \frac{d\Gamma_{\tau^+}}{d\omega} \right) d\omega}{\frac{1}{2} \iiint_{Q^2_{1,i}}^{Q^2_{2,i}}  \left(\frac{d\Gamma_{\tau^-}}{d\omega}+\frac{d\Gamma_{\tau^+}}{d\omega} \right)d\omega}  \nonumber \\
&\simeq \langle \cos\beta\cos\psi \rangle^i_{\tau^-} - \langle \cos\beta\cos\psi \rangle^i_{\tau^+} 
\label{Eq:ACP_def}
\end{align}
with $d\omega=dQ^2 d\!\cos\theta d\!\cos\beta$. In other words, $\ACPth$ is the difference between the mean values of $\cos\beta\cos\psi$ for $\tau^+$ and $\tau^-$ events evaluated in bins of $Q^2$.

We use $699\,\mathrm{fb}^{-1}$ of data collected at the $\Upsilon(3S)$, $\Upsilon(4S)$ and $\Upsilon(5S)$ resonances and off-resonance with the Belle detector~\cite{Abashian:2000cg} at the KEKB asymmetric-energy $e^+e^-$ collider \cite{Kurokawa:2001nw}.  
The signal and backgrounds from $\tau^+\tau^-$ events are generated by KKMC/TAUOLA \cite{Jadach:1999vf}. 
The detector response is simulated by a GEANT3 \cite{Brun:1987ma} based program. 


Using standard event topology requirements, a $e^+e^-\rightarrow \tau^+\tau^-(\gamma)$ sample is selected as described in \cite{Fujikawa:2008ma}.

In the CM frame, the event is divided into two hemispheres using the plane perpendicular to the direction of the thrust axis \cite{Farhi:1977sg}. 
Events with one charged track from an electron, muon or pion in one hemisphere (tag side) and a charged pion and a $K^0_S\rightarrow \pi^+\pi^-$ candidate in the other hemisphere (signal side) are chosen. The $K^0_S$ candidates are required to have an invariant mass in the range $0.485\,\mathrm{GeV}/c^2 <M_{\pi\pi} < 0.511\,\mathrm{GeV}/c^2$ and a reconstructed $K^0_S$ decay length greater than $2\,\mathrm{cm}$. 
The selection criteria for the signal side and particle identification criteria are described in detail in \cite{Epifanov:2007rf}. Backgrounds from decays with a $\pi^0$ are suppressed by rejecting events containing photons on the signal side with energies greater than $0.15\,\mathrm{GeV}$.  To further suppress background from $e^+e^-\rightarrow q\bar{q}$ ($q=u,d,s,\;\mathrm{and}\;c$) processes, a thrust value above $0.9$ is required and for events with a pion on the tag side, the number of tag side photons with energies greater than $0.1\,\mathrm{GeV}$  must be less than five.
In total,  $(162.2 \pm 0.4)\times 10^3$ $\tauToKSpiP$  and  $(162.0 \pm 0.4)\times 10^3$ $\tauToKSpiM$ candidates are selected.
Background contributions from $\tau$ decays with the exception of $\tau^\pm \rightarrow \nu_\tau \pi^\pm \pi^+ \pi^-$ and contributions from $e^+e^-\rightarrow {q}\bar{q}$ and two-photon processes are estimated from Monte Carlo (MC) simulation \cite{Lange:2001uf,Jadach:1991by, Berends:1986ig} using the branching fractions from  \cite{Nakamura:2010pdg}.  Contributions from $\tau^\pm \rightarrow \nu_\tau \pi^\pm \pi^+ \pi^-$ are estimated using the data in the two $K^0_S$ sideband regions, $0.469\,\mathrm{GeV}/c^2 <M_{\pi\pi} < 0.482\,\mathrm{GeV}/c^2$ and $0.514\,\mathrm{GeV}/c^2 <M_{\pi\pi} < 0.527\,\mathrm{GeV}/c^2$ 
\footnote{The contributions of the simulated background modes were subtracted from the data in $K^0_S$ sideband regions in order to avoid double counting.}. 

\begin{figure}[!htpb] 
\begin{center}
\includegraphics[width = 70mm,]{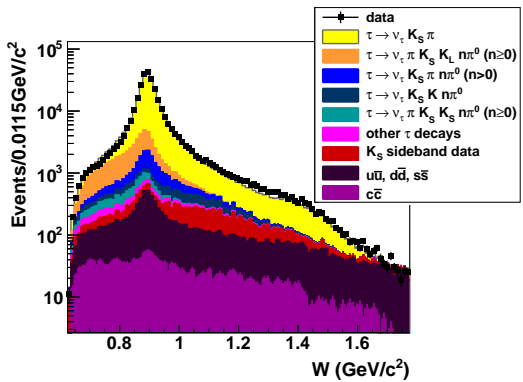}
\caption[]{Mass spectrum of the $K^0_S\pi^\pm$ system. Data are indicated by the squares, simulated signal and the estimated background contributions are shown by the colored histograms. All background modes have been determined from Monte Carlo with exception of $\tau^\pm \rightarrow \nu_\tau \pi^\pm \pi^+ \pi^-$ which has been estimated from $K^0_S$ sideband data.
}
\label{fig:Whad}
\end{center}  
\end{figure}

The largest background contribution is due to other $\tau$ decays, namely $(9.5 \pm 3.2)\%$ of the events in the selected signal sample from $\tau^\pm \rightarrow \nu_\tau K^0_SK^0_L \pi^\pm$,  $(3.7\pm 1.2)\%$ from $\tau^\pm \rightarrow \nu_\tau K^0_S \pi^\pm \pi^0$,  $(1.7\pm0.2)\%$ from $\tau^\pm \rightarrow \nu_\tau K^0_S K^\pm$, and  $(1.79\pm0.03)\%$ from $\tau^\pm \rightarrow \nu_\tau \pi^\pm \pi^+ \pi^-$.
The contribution from $e^+e^-\rightarrow q\bar{q}$ is $(3.4\pm1.0)\%$. The backgrounds from $b\bar{b}$, Bhabha and two-photon processes are negligible. The total contribution of background processes is $(22.1\pm3.6)\%$. 
The invariant mass of the $K^0_S\pi^\pm$ system, $W=\sqrt{Q^2}$, for the selected data events is shown in Fig.~\ref{fig:Whad} together with simulated signal events and the background contributions discussed above. Signal events were generated by a modified version of TAUOLA that incorporates the results of  \cite{Epifanov:2007rf}.

To avoid possible bias, the CPV search is performed as a blind analysis. First, possible sources of artificial CPV, such as forward-backward (FB) asymmetries in the $e^+e^-\rightarrow\tau^+\tau^-$ production ($\gamma-Z$ interference effects and higher-order QED effects) and detector induced differences between $\pi^+$ and $\pi^-$ reconstruction efficiencies, are studied using data.  Other unknown sources are investigated in data by measuring the CP asymmetry in a control sample  described below. 

The FB asymmetry is measured in $\tau^\pm\rightarrow \nu_\tau\pi^\pm\pi^+\pi^-$ events (excluding $K_S^0\rightarrow \pi^+\pi^-$ signal candidates by using a mass and decay length veto) as a function of the momentum and polar angle of the $\pi^\pm\pi^+\pi^-$ system. 
An effect of a few percent is observed, which is described well by the MC simulation. The asymmetry for $\pi^\pm$ detection,  which can arise because of the different nuclear interaction cross sections for positively and negatively charged hadrons,  is studied in the laboratory system as a function of momentum and polar angle of the charged pions in $\tau^\pm\rightarrow \nu_\tau\pi^\pm\pi^+\pi^-$ (excluding $\pi^+\pi^-$ combinations consistent with $K^0_S$ decays) events and found to be of O($10^{-2}$) (see \cite{bischofberger:2011EPAPS} for details).
Using these measurements, correction tables are obtained that are then applied as weights for each event. Since the CP asymmetry is measured as a function of angles relative to the $\tau$ direction rather than polar angles in the laboratory, the net effect of these corrections on the CP asymmetry is very small [O($10^{-4}$) for FB asymmetry effects and O($10^{-3}$) for the $\pi^\pm$ detection asymmetry].  
   
A control sample is selected from $\tau^\pm\rightarrow \nu_\tau\pi^\pm\pi^+\pi^-$ events \footnote{Possible CP violation in this decay mode is expected to be small because of the small Higgs coupling to the $d$ quark and is smeared out if only the two-body decay angle $\beta$ (and $\psi$) is measured (see \cite{Kiers:2008mv}).} by requiring that the invariant mass of both $\pi^+\pi^-$ combinations lie outside of the $K^0_S$ mass window but the mass of one of the combinations lie in the sideband of this window. The resulting sample consists of about $10^6$ events, i.e. about three times more than the signal sample.
The CP asymmetry measured in this control sample is very small [O($10^{-3}$)] (see \cite{bischofberger:2011EPAPS} for details) and serves as an estimate of the remaining unknown systematic effects.
                                                                                                        
The observed CP asymmetry in the selected $\tauToKSpi$ candidate sample is shown in Table~\ref{table::Asymmetry_SubDataWFBWDT} for four bins of the hadronic mass $W = \sqrt{Q^2}$ before and after applying the corrections for 
higher-order QED and $\pi^\pm$ detection asymmetry effects. 
\begingroup
\squeezetable
\begin{table} 
\caption[]{CP asymmetry $\ACPsig$ measured in bins of the hadronic mass $W$.  
The 2nd and 3rd column show the observed asymmetry with statistical errors only, before and after correcting for higher-order QED and $\pi^\pm$ detection asymmetry effects. The final CP asymmetry after background subtraction is shown in the 4th column where first and second errors correspond to statistical and systematic errors, respectively. The 5th column shows the observed number of signal events $n_i$ per $W$ bin (after background subtraction) divided by $N_s=\sum_i n_i$.
\label{table::Asymmetry_SubDataWFBWDT}}
\begin{ruledtabular}
\begin{tabular}{ l   c   c   c   c  } 
\multicolumn{1}{c}{$W$}& \multicolumn{3}{c}{$\ACPsig$  ($10^{-3}$)}  &\\  
\multicolumn{1}{c}{($\mathrm{GeV}/{c^2})$}  & Observed & Corrected & Backgr. subtr. &  $n_i/N_s (\%)$ \\ \hline 
$0.625\! -\! 0.890$      & $-0.1	\pm	2.1$   & $5.2  \pm 2.1$  &             $7.9 \pm 3.0  \pm  2.8$    & $36.53  \pm  0.14 $ \\ 
$0.890\! -\! 1.110$      & $-2.7	\pm	1.7$   & $1.6  \pm 1.7$  &             $1.8 \pm 2.1  \pm  1.4$   & $57.85  \pm 0.15$   \\  
$1.110\! -\! 1.420$      & $-5.1	\pm	4.7$   & $-3.5 \pm 4.7$  &            $-4.6 \pm 7.2  \pm  1.7$    & $4.87   \pm 0.04$    \\ 
$1.420\! -\! 1.775$      & $9.3	\pm	12.1$ & $9.6  \pm 12.1$ &           $-2.3 \pm 19.1\pm  5.5$ & $0.75 \pm 0.02$  \\   
\end{tabular}
\end{ruledtabular}
\end{table}
\endgroup
The 4th column shows the final values of the CP asymmetry after subtraction of the background contributions.
Here, we assume that there is no CP asymmetry in the background and correct the background effects as 
\begin{equation}
\ACPthi = \frac{\langle \cos\beta\cos\psi \rangle^i_{\tau^-}}{1-f^-_{b,i}} - \frac{\langle \cos\beta\cos\psi \rangle^i_{\tau^+}}{1-f^+_{b,i}}
\end{equation}
where $f^\pm_{b,i}$ are the fractions of background in the selected $\tau^\pm$ samples in $W$ bin $i$.

In order to account for possible systematic uncertainties due to detector effects, the quadratic sum of the values of $\ACPsig$ measured in the control sample and their statistical errors are used as an estimate of the systematic error. Other contributions to the systematic error arise in the background subtraction because of uncertainties in the estimated number of background candidates and limited MC statistics. These contributions are however small in comparison. A summary of the systematic uncertainties is given in Table~\ref{table::Asymmetry_SubDataWFBWDT_Err}.  
\begingroup
\squeezetable
\begin{table} 
\caption[]{Systematic uncertainties in the CP asymmetry $\ACPsig$. The second column shows the uncertainties due to effects introduced by the detector, which are estimated from the $\ACPsig$ measurement in the control sample. Contributions from uncertainties in the background estimates and limited MC statistics are small in comparison.
\label{table::Asymmetry_SubDataWFBWDT_Err}} 
\begin{ruledtabular}
\begin{tabular}{ l  c  c c  c }
\multicolumn{1}{c}{$W$}                                   &\multicolumn{4}{c}{Systematic uncertainties ($10^{-3}$) }    \\
\multicolumn{1}{c}{($\mathrm{GeV}/{c^2})$}   &Detector    & Backgr.  & MC stat.  &Total     \\ \hline
$0.625\! -\! 0.890$                                           & 2.76         &0.59        & 0.15        & 2.83    \\ 
$0.890\! -\! 1.110$                                           & 1.40         & 0.04       & 0.10        &  1.40  \\  
$1.110\! -\! 1.420$                                           & 1.50         & 0.25       & 0.79        &  1.71    \\
$1.420\! -\! 1.775$                                           & 5.18         & 0.96       & 1.38        &  5.45  \\   
\end{tabular}
\end{ruledtabular}
\end{table}
\endgroup

\begin{figure}[!htpb] 
\begin{tabular}{cc}
   \includegraphics[width = 42mm]{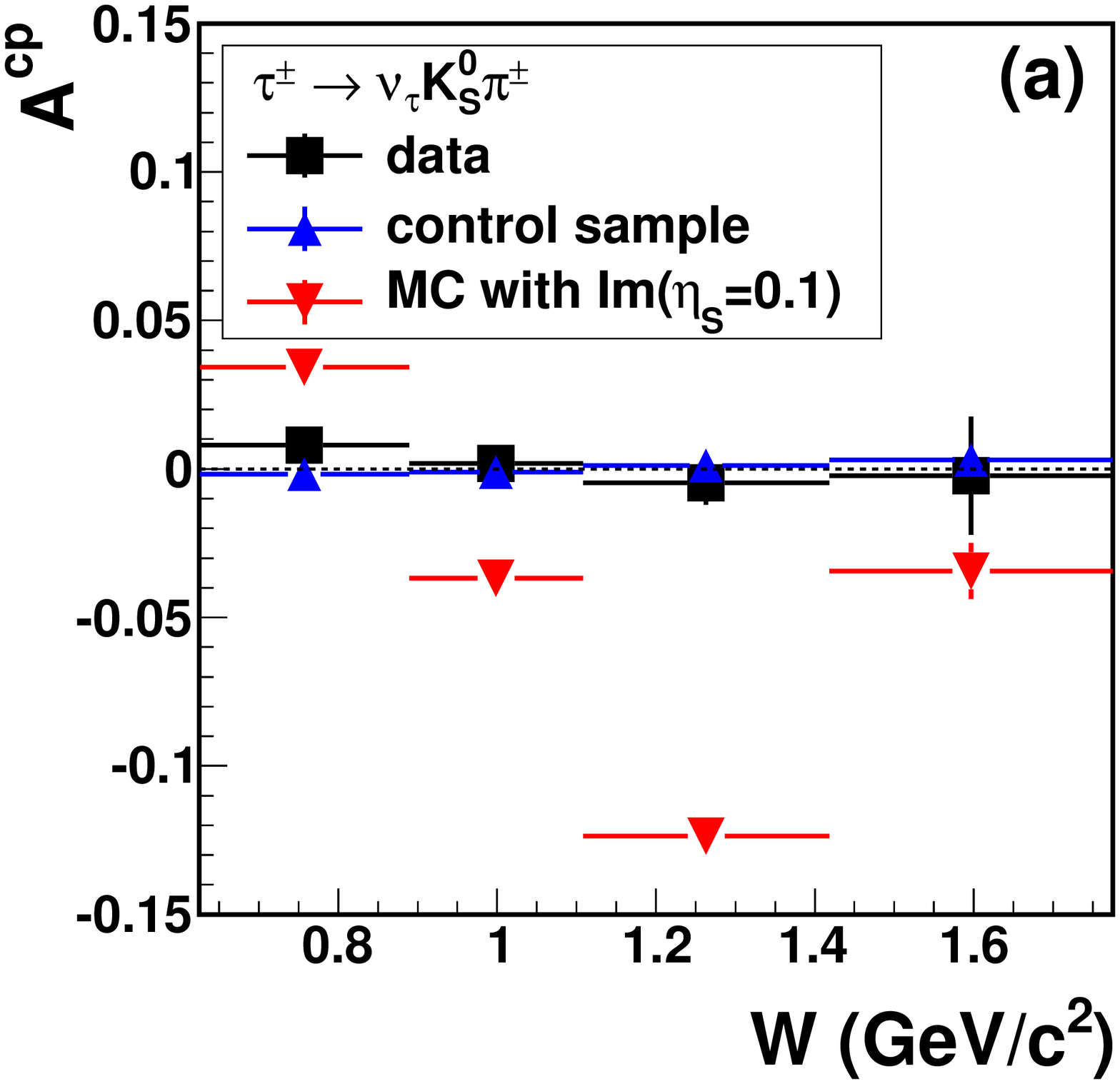} &
   \includegraphics[width = 42mm]{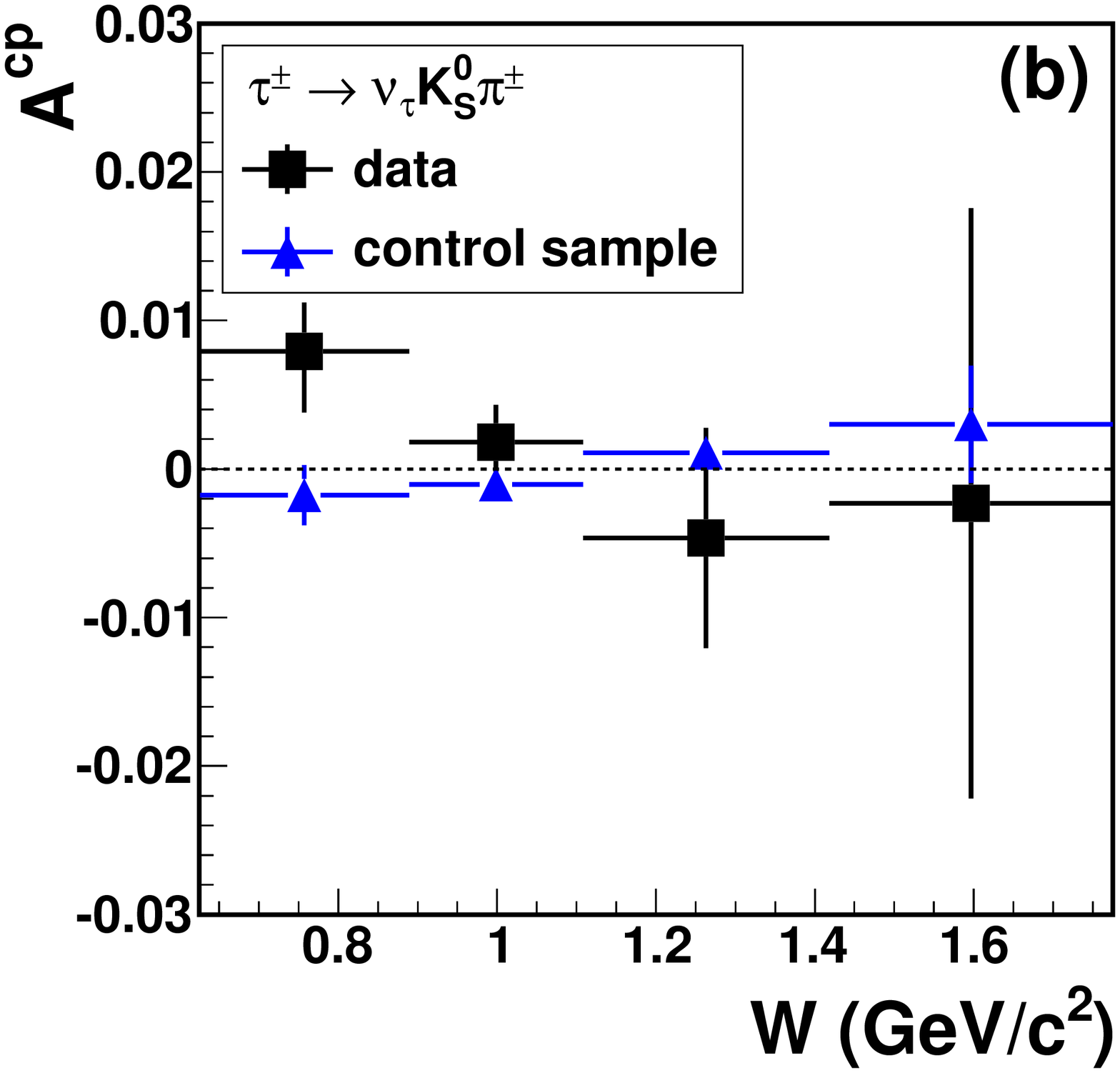}
\end{tabular}
\caption[]{(a) Measured CP violation asymmetry after background subtraction (squares). The vertical error bars are the statistical error and systematic errors added in quadrature. The CP asymmetry measured in the control sample is indicated by the blue triangles (statistical errors only) and the inverted red triangles show the expected asymmetry for $\ImETA = 0.1$ [$\ReETA =0$]. (b) Expanded view (the vertical scale is reduced by a factor of five). \label{fig:AsymmetryBetaPsi_Final}}
\end{figure}

The background subtracted asymmetry is shown in Fig.~\ref{fig:AsymmetryBetaPsi_Final} (a) and (b) with statistical and systematic errors added in quadrature. The asymmetry is small and except for the lowest mass bin within one standard deviation ($\sigma$) of zero.
For comparison the predicted CP asymmetry is shown in Fig.~\ref{fig:AsymmetryBetaPsi_Final} (a) for $\ImETA = 0.1$ and $\ReETA =0$ \footnote{The prediction is obtained from MC by using {\it solution 1} of Table 4 in \cite{Epifanov:2007rf} as parameterizations for the form factors $F$ and $F_S$ together with Eq.~(\ref{eqn:FHfromFS}).}. Note that the current best limit by the CLEO experiment \cite{Bonvicini:2001xz} corresponds to $|\ImETA|<0.19$.

From the measured values of $\ACPsig$ the CPV parameter $\ImETA$ can be extracted, which allows an interpretation in the context of NP models. Taking into account the detector efficiencies, the relation between $\ACPsig$ and $\ImETA$ is given as
 \begin{align}
\ACPsigi  \simeq  \ImETA   \frac{N_s}{n_i} \int_{Q^2_{1,i}}^{Q^2_{2,i}} C(Q^2) \frac{\IMcs(FF_H^*)}{m_\tau}\,dQ^2 \label{eqn:AymmetryBetaPsi_mod} 
  \equiv  c_i \ImETA, 
\end{align}
where $n_i$ is the observed number of $\tauToKSpi$ events in $Q^2$ bin $i$ $(Q^2 \in [Q^2_{1,i},Q^2_{2,i}])$ and $N_s=\sum_i n_i$ is the total number of observed $\tauToKSpi$ events. The function $C(Q^2)$ includes the detector efficiency as well as all model-independent terms. First, the efficiency is determined as a function of $Q^2$,  $\beta$ and $\theta$, then $C(Q^2)$ is obtained after numerical integration over the decay angles $\beta$ and $\theta$. The parameterization of $C(Q^2)$ is given in \cite{bischofberger:2011EPAPS}.

Using the function $C(Q^2)$ and the fractions ${N_s}/{n_i}$ which are given in Table~\ref{table::Asymmetry_SubDataWFBWDT}, the linearity constants $c_i$, which relate $\ACPsig$ and $\ImETA$, can be determined for any parameterization of the form factors $F$ and $F_H$ simply by calculating the integral in Eq.~(\ref{eqn:AymmetryBetaPsi_mod}) \footnote{This allows for a simple re-evaluation of limits for $\ImETA$ when better knowledge for the form factors is available from theory or future measurements.}.  

To determine limits for $|\ImETA |$, three parameterizations of $F$ and $F_S$ [exploiting Eq.~(\ref{eqn:FHfromFS})] as linear combinations of Breit-Wigner shapes of the vector resonances $K^*(892)$ and $K^*(1410)$ and the scalar resonances $K^*_0(800)$ and $K^*_0(1430)$ are used. These parameterizations were determined in an earlier Belle measurement of the $K^0_S\pi^\pm$ mass spectrum \cite{Epifanov:2007rf}.
In addition, a constant strong interaction phase difference between $F$ and $F_S$, $\phi_S = \mathrm{arg}[F_S(Q^2_\mathrm{min})] - \mathrm{arg}[F(Q^2_\mathrm{min})]$ with $Q^2_\mathrm{min} = (m_\pi +m_{K^0_S})^2$, is introduced for generality because such a relative phase cannot be determined from the $K^0_S\pi^\pm$  mass spectrum.

Using Eq.~(\ref{eqn:AymmetryBetaPsi_mod}), the linearity constants $c_i$ are calculated in each mass bin for $\phi_S = 0^\circ, 5^\circ,\ldots ,360^\circ$ and the obtained values of $\ImETA$ with associated uncertainties are combined to determine upper limits for $| \ImETA |$.  
For each parameterization, the value $\phi_S$ giving the most conservative limit is chosen. For the three parameterizations of $F$ and $F_S$, this results in the range of limits  $|\ImETA| < (0.012 - 0.026)$ at $90\%$ confidence level. If we fix $\phi_S\equiv0$, the range $|\ImETA| < (0.011 - 0.023)$ is obtained.
The parameterizations of $F$ and $F_S$ used by the CLEO collaboration \cite{Bonvicini:2001xz} yield a comparable limit  $|\ImETA| < 0.013$.
These results are about one order of magnitude more restrictive than the previous best upper limit, $|\ImETA| < 0.19$, obtained by the CLEO collaboration \cite{Bonvicini:2001xz}.

Theoretical predictions for $\ImETA $ can be given in context of a MHDM with three or more Higgs doublets \cite{Weinberg:1976hu,Grossman:1994jb}. 
In such models $\eta_S$ is given by \cite{Choi:1994ch}
\begin{equation}
\eta_S \simeq  \frac{m_\tau m_s}{M_{H^\pm}^2} X^*Z   \label{eq:etaSUSY}
\end{equation}
if numerically small terms proportional to $m_u$ are ignored. 
Here, $M_{H^\pm}$ is the mass of the lightest charged Higgs boson and the complex constants $Z$ and $X$ describe the coupling of the Higgs boson to the $\tau$ and $\nu_{\tau}$ and the $u$ and $s$ quarks, respectively (see \cite{Grossman:1994jb,Choi:1994ch}). The limit $|\ImETA |<0.026$ is therefore equivalent to 
\begin{equation}
|\IMcs(XZ^*)|<0.15  \frac{M_{H^\pm}^2} {1\,\mathrm{GeV}^{2}/c^4}.
\end{equation}


In summary, we have searched for CP violation in $\tauToKSpi$ decays, analyzing the decay angular distributions. No significant CP asymmetry has been observed. Upper limits for the CP violation parameter $\ImETA$ at $90\%$ confidence level are in the range $|\ImETA |<0.026$ or better, depending on the parameterization used to describe the hadronic form factors and improve upon previous limits by one order of magnitude.


\begin{acknowledgments}
We acknowledge J.~H.~K\"uhn, K.~Kiers and  T.~Morozumi for useful discussions.
We thank the KEKB group for excellent operation of the
accelerator, the KEK cryogenics group for efficient solenoid
operations, and the KEK computer group and
the NII for valuable computing and SINET4 network support.  
We acknowledge support from MEXT, JSPS and Nagoya's TLPRC (Japan);
ARC and DIISR (Australia); NSFC (China); MSMT (Czechia);
DST (India); MEST, NRF, NSDC of KISTI, and WCU (Korea); MNiSW (Poland); 
MES and RFAAE (Russia); ARRS (Slovenia); SNSF (Switzerland); 
NSC and MOE (Taiwan); and DOE (USA).
\end{acknowledgments}


%
\end{document}


\newcommand{\tauToKSpi} {\tau^\pm  \rightarrow     K^0_S        \pi^\pm      \nu_\tau}
\newcommand{\tauToKSpiM}{\tau^-     \rightarrow     K^0_S        \pi^-    \nu_\tau}
\newcommand{\tauToKSpiP}{\tau^+     \rightarrow     K^0_S        \pi^+    \bar{\nu}_\tau}

\newcommand{\ACPth}{A^\text{cp}}
\newcommand{\ACPsig}{A^\text{cp}}
\newcommand{\ACPsigi}{A^\text{cp}_{i} }

\newcommand{\ReETA}{\Re(\eta_S)}
\newcommand{\ImETA}{\Im(\eta_S)}
\newcommand{\IMcs}{\Im}



\title{Supplementary Material}

\maketitle

{\renewcommand{\thefootnote}{\fnsymbol{footnote}}}
\setcounter{footnote}{0}
\section{Derivation of correction factors for experimental asymmetries}
Possible sources of artificial CP violation (CPV) are the forward-backward (FB) asymmetry in the  $e^+e^-\rightarrow\tau^+\tau^-$ production and
 the asymmetry for the $\pi^{\pm}$ detection. For both of these experimental asymmetries, correction factors have been prepared using data.  

\subsection{Forward-backward asymmetry}
\begin{figure}[!htpb] 
\begin{center}
\includegraphics[height = 50mm]{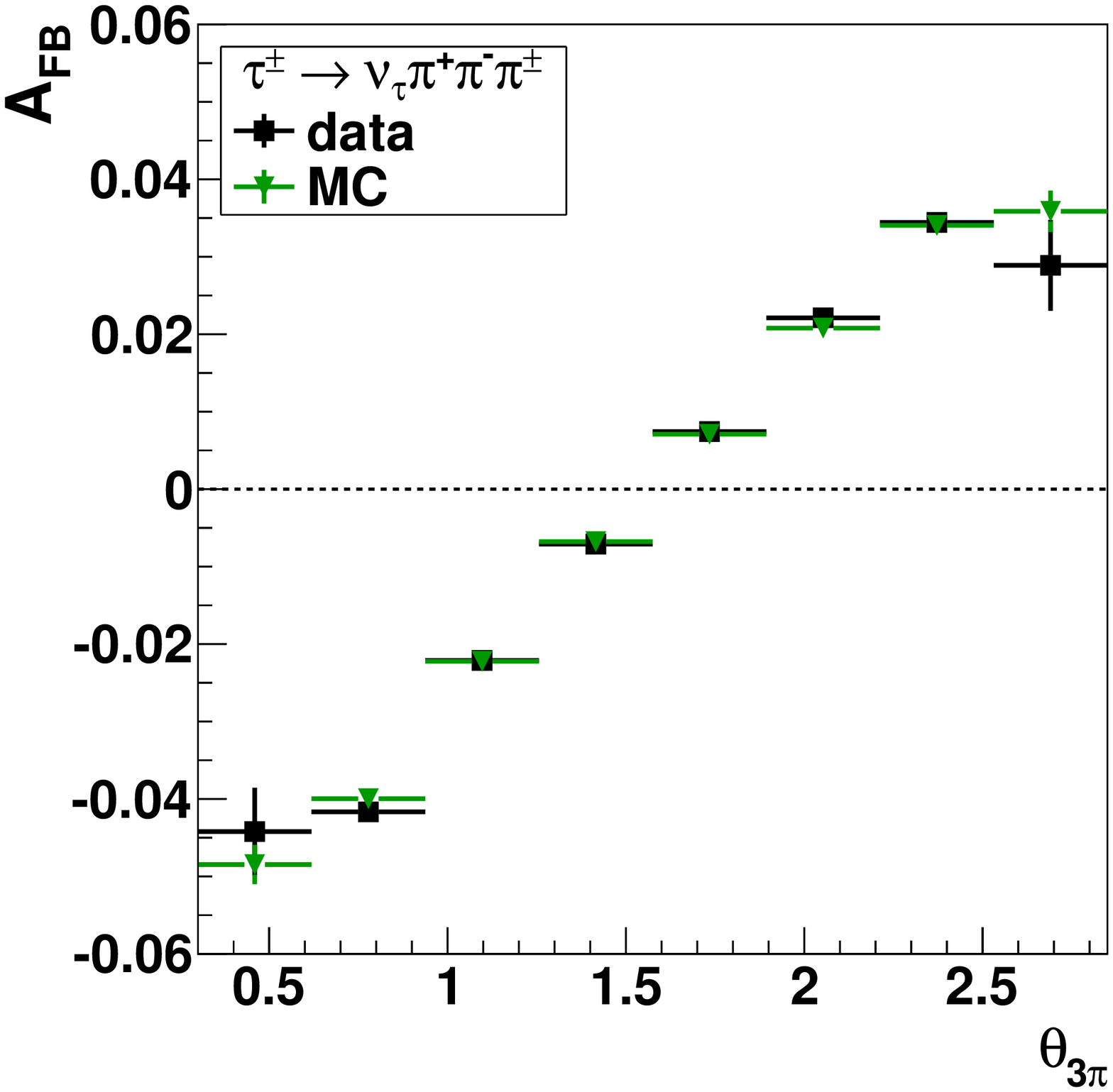}
\caption[]{ Forward-backward asymmetry as a function of the polar angle of the three pions, $\theta_{3\pi}$, in the CM frame (z-axis along the direction opposite to the positron beam)
for $\tau^{+}\tau^{-}$ events with $\tau^{\pm}\rightarrow\nu_{\tau} \pi^{\pm}\pi^{+}\pi^{-}$ on one side and a one prong decay on the other side. 
The asymmetry is shown for data and Monte Carlo (MC).}
\label{fig:FB}
\end{center}  
\end{figure}

The FB asymmetry is  measured  
using $\tau^+\tau^-$ events with  $\tau^{\pm}\rightarrow\nu_{\tau} \pi^{\pm}\pi^{+}\pi^{-}$ 
on one side and a one prong decay on the other side.  Using the direction of $\pi^{\pm}\pi^{+}\pi^{-}$ system  as an approximation of the direction of the $\tau$ lepton, the
measured FB asymmetry is found to be a few percent as shown in Fig.~\ref{fig:FB} and well described by Monte Carlo.
In order to take this FB asymmetry into account, a correction weight factor is prepared as a function of the momentum and polar angle of the $\pi^{\pm}\pi^{+}\pi^{-}$  system
and is applied for each event.
The net effect of this correction on the measured CP asymmetry is very small [O($10^{-4}$)].

\subsection{Asymmetry for the $\pi^{\pm}$ detection}
\begin{figure}[!htpb] 
\begin{center}
\includegraphics[width = 50mm,]{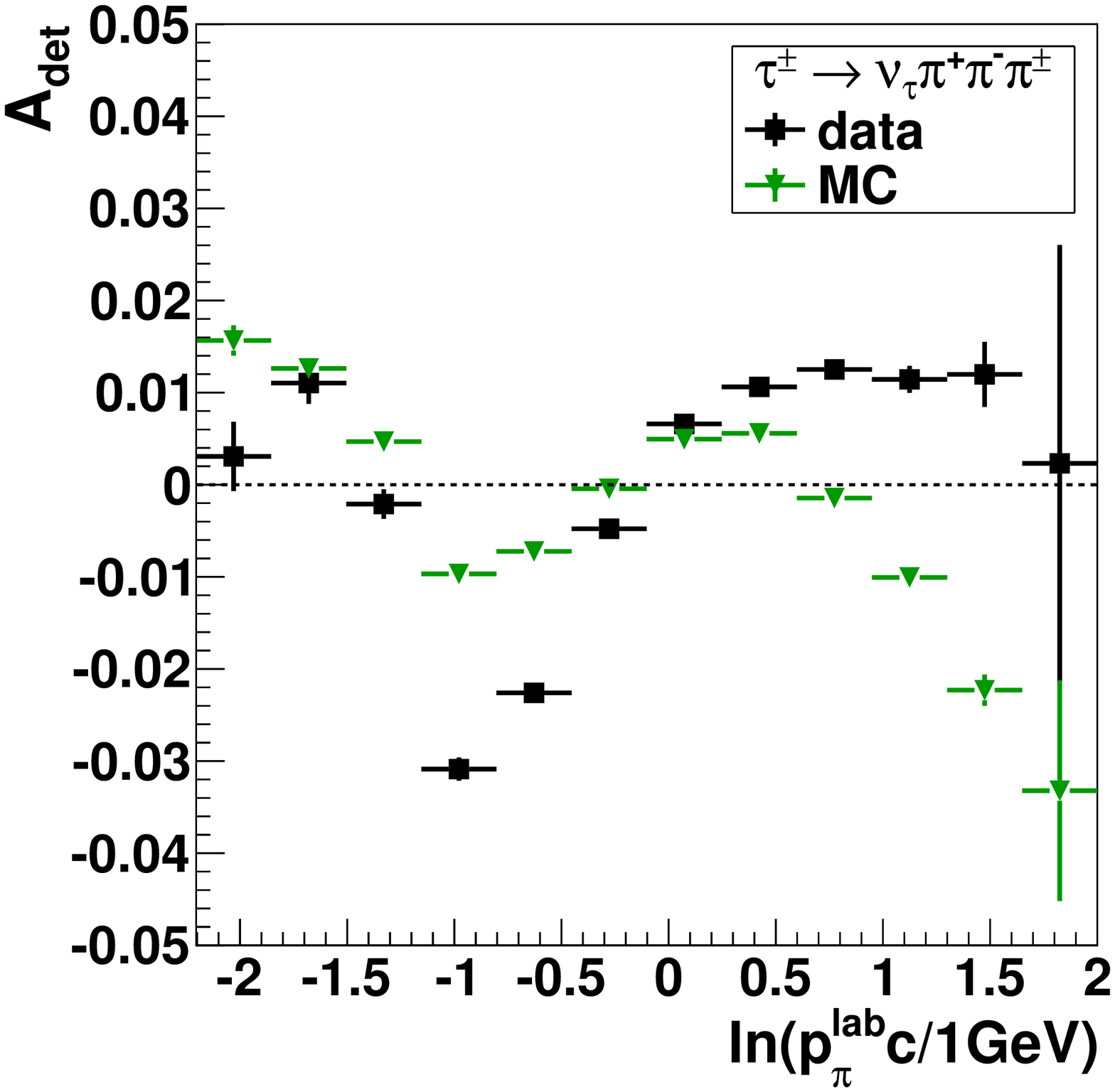}
\caption[]{The $\pi^{\pm}$ detection asymmetry as a function of the pion momentum $\ln ( p_{\pi}^{\text{lab}}\text{c}/1\,\text{GeV})$ in the
laboratory frame for data and Monte Carlo (MC). The asymmetry is shown after applying the correction for the forward-backward asymmetry. 
A logarithmic scale is chosen for the momentum in order to obtain a more even distribution of events. The current MC gives a poor description of this asymmetry because the nuclear cross sections for $\pi^\pm$ are not well known.
} 
\label{fig:piasym}
\end{center}  
\end{figure}

The different nuclear interaction cross section for
$\pi^{+}$ and $\pi^{-}$ can lead to a difference in tracking and particle identification efficiencies
depending on the polar angle and momentum of the pion in the laboratory frame.
This detector induced asymmetry is measured using the same data sample as for the FB asymmetry
studies. The charge asymmetry
\begin{equation*}
A_{\rm det} ( \theta_ {\pi}, p_{\pi} ) = \frac{ n^{+}   - n^{-}} { n^{+} + n^{-}}
\end{equation*}
is measured as a function of 
the polar angle $\theta_{\pi}$ and the momentum $p_{\pi}$ of $\pi^{\pm}$ in the laboratory system,
where $n^{\pm}$ denotes the number of $\pi^{\pm}$ particles in each $\theta_{\pi}$ and $p_{\pi}$
bin.  The observed $\pi^{\pm}$ detection asymmetry is up to a few percent as shown  in Fig.~\ref{fig:piasym} and
taken into account by applying correction weight factors
\begin{equation*}
  \omega^{\pm}_{\rm det} (\theta_{\pi}, p_{\pi} ) = \frac{ n^{+} + n^{-}} {2 n^{\pm}}
\end{equation*}
for each events as a function of the charge, the momentum and the polar angle of the $\pi^{\pm}$ 
in the signal $\tauToKSpi$ events.
The net effect of this correction on the measured CP asymmetry is of the order of  [O($10^{-3}$)] as shown in Table I of this publication.

\section{CP asymmetry in the control sample}

In order to check for any remaining detector introduced bias after the corrections for the FB asymmetry and the $\pi^{\pm}$ detection asymmetry,
the CP asymmetry is measured using the data in the control sample. The measured CP asymmetry in this sample before and after applying the 
corrections weights as well as the number of the events in the control sample are given inTable~ \ref{table::Asymmetry_Ctr}.
\begingroup
\begin{table} [h]
\caption[]{CP asymmetry $\ACPsig$ measured in the control sample for four bins of the hadronic mass $W$. The 2nd and 3rd column show the observed asymmetry with statistical errors only, before and after correcting for the FB asymmetry and $\pi^\pm$ detection asymmetry effects. The 5th column shows the number of selected events in the control sample.  
\label{table::Asymmetry_Ctr}}
\begin{ruledtabular}
\begin{tabular}{ l   c   c   c   c  } 
\multicolumn{1}{c}{$W$}&  \multicolumn{2}{c}{$\ACPsig$  ($10^{-3}$)} & Number of events\\  
\multicolumn{1}{c}{($\mathrm{GeV}/{c^2})$} &   Observed & Corrected & $n_i$ \\ \hline 
$0.625\! -\! 0.890$     & $-7.3 \pm 2.0$   &  $-1.8 \pm 2.0$ &   $129993 \pm 361$  \\ 
$0.890\! -\! 1.110$     &  $-4.4 \pm 0.9$  &  $-1.0 \pm 0.9$ &   $529559 \pm 728$  \\  
$1.110\! -\! 1.420$     &  $-0.3 \pm 1.0$  &  $ 1.1 \pm 1.0$ &   $367181 \pm 606$   \\ 
$1.420\! -\! 1.775$     &  $3.2  \pm 3.9$  &  $ 3.0 \pm 3.9$ &   $20274   \pm 142$   \\   
\end{tabular}
\end{ruledtabular}
\end{table}
\endgroup

\section{ Parametrization of $C(Q^{2})$ in Eq.~(9)}

The measured CPV asymmetry $\ACPsig$ can be related to the CPV parameter $\ImETA$ by
 \begin{align}
\ACPsigi  &= \langle \cos\beta\cos\psi \rangle^i_{\tau^-} - \langle \cos\beta\cos\psi \rangle^i_{\tau^+} \\
& \simeq  \frac{1}{n_i}\left( {\sum_{j\in \tau^-}}^i\cos\beta_j\cos\psi_j - {\sum_{k\in \tau^+}}^i\cos\beta_k\cos\psi_k \right) \nonumber\\
& =  \frac{N_s}{n_i} \frac{1}{N_s}  \left( {\sum_{j\in \tau^-}}^i\cos\beta_j\cos\psi_j - {\sum_{j\in \tau^+}}^i\cos\beta_j\cos\psi_j \right) \nonumber \\
& =  \frac{N_s}{n_i} \frac{1}{\epsilon_\text{tot} \Gamma } \iiint_{Q^2_{1,i}}^{Q^2_{2,i}} \epsilon(Q^2,\cos\beta, \cos\theta)  \nonumber \\
& \times \cos\beta \cos\psi \left[ \frac{d\Gamma(\tau^-)}{d\omega} - \frac{d\Gamma(\tau^+)}{d\omega} \right] \, d\omega \\ \nonumber 
&\simeq  \ImETA   \frac{N_s}{n_i} \int_{Q^2_{1,i}}^{Q^2_{2,i}} C(Q^2) \frac{\IMcs(FF_H^*)}{m_\tau}\,dQ^2 \label{eqn:AymmetryBetaPsi_mod}, 
\end{align}
with $d\omega=dQ^2 d\!\cos\theta d\!\cos\beta$. Here $n_i = (n_i^- + n_i^+)/2$, where $n_i^\pm$ denotes the observed number of
 $\tauToKSpi$ events in $Q^2$ bin $i$ $(Q^2 \in [Q^2_{1,i},Q^2_{2,i}])$ and $N_s=\sum_i n_i$. 
The coefficients $\epsilon_\text{tot}$ and $\epsilon(Q^2,\cos\beta,\cos\psi)$ are the total and the three dimensional detector efficiencies and $\Gamma$ is the total $\tauToKSpi$ decay width. The function $C(Q^2)$ contains the model independent terms and detector efficiency effects and  is obtained after numerical integration over $\cos\beta$ and $\cos\theta$:
\begin{align}
C(Q^2) & = -\frac{1}{\Gamma} \frac{G^{2}_{F}}{2m_{\tau}}\sin^{2}\theta_c 
  \frac{1}{(4\pi)^3} \frac{(m_\tau^2-Q^2)^2}{Q^2}|\vec{q}_1|^{2}  \\
              & \times  \iint   \frac{\epsilon(Q^2,\cos\beta,\cos\psi)}{\epsilon_\text{tot}}  \cos^2\beta \cos^2\psi  \, d\!\cos\theta \,d\!\cos\beta. \nonumber
\end{align}
For a fixed value of $Q^2$, the relation between $\cos\theta$ and $\cos\psi$ can be obtained from Eq.~(2) and Eq.~(5) of this publication.

\begin{figure}[!htpb] 
\begin{center}
\includegraphics[width = 70mm,]{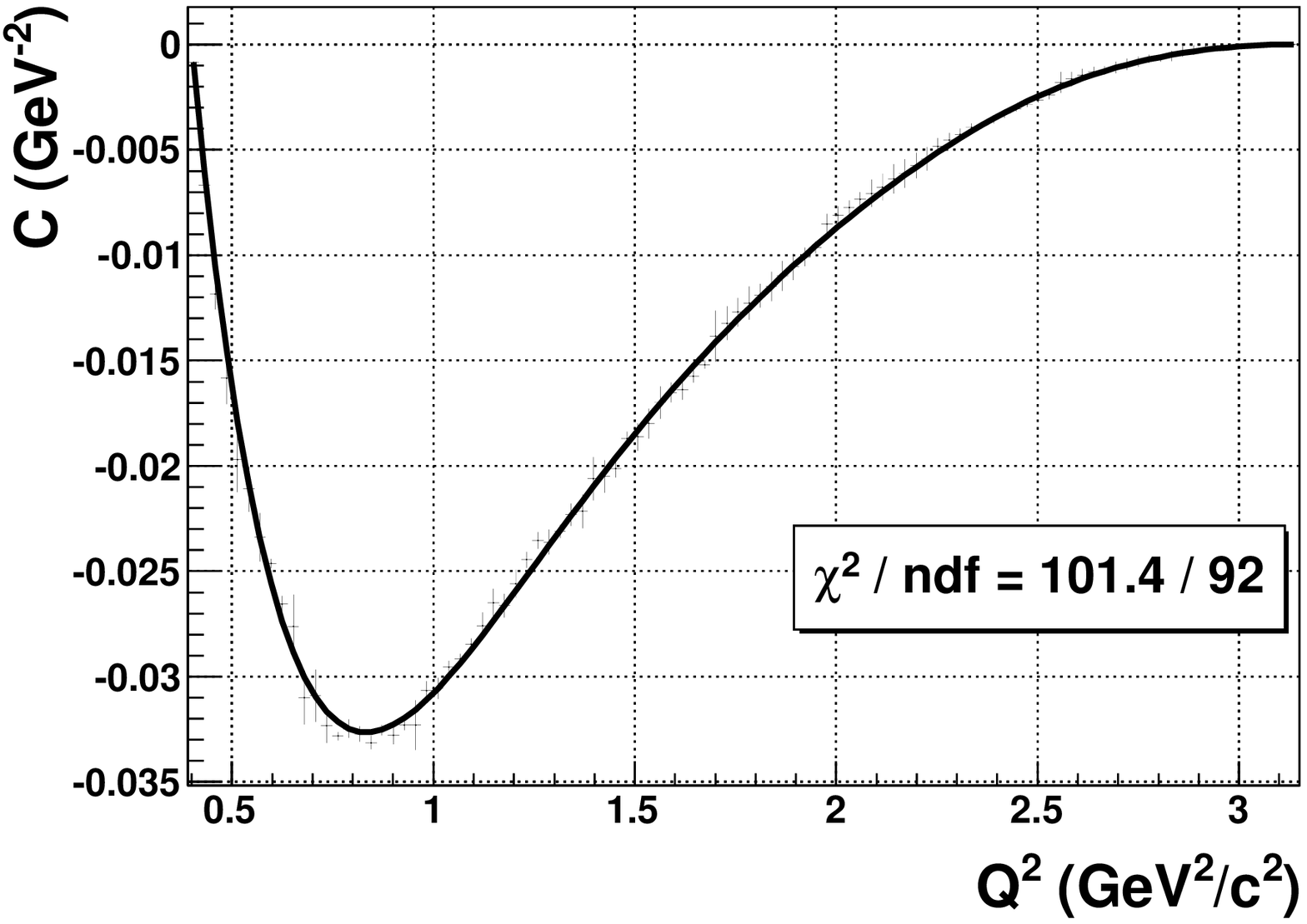}
\caption[]{Fit of the function $C$ which is used to related the measured CPV asymmetry $\ACPsig$ to the CP violation parameter $\ImETA$. The function is obtained after numerical integration over the cosine of the decay angles $\beta$ and $\theta$ and includes detector efficiency effects. The goodness of the fit is $\chi^2/\text{ndf} = 101.4/92$ where $\mathrm{ndf}$ denotes the number of degrees of freedom.}
\label{fig:fittedC}
\end{center}  
\end{figure}

The resulting $C(Q^2)$ can be parameterized as a 7th-order polynomial,
\begin{equation}
C({Q}^{2}) = \sum_{i=0}^{7} a_i  \bar{Q}^{2i}
\end{equation} 
where $\bar{Q}^2$ denotes the dimensionless value of $Q^2$ measured in units of $\mathrm{GeV}^2/c^4$. The obtained fit result is shown in Fig~(\ref{fig:fittedC}). The goodness of the fit is $\chi^2/\text{ndf} = 101.4/92$ where $\mathrm{ndf}$ denotes the number of degrees of freedom.
The uncertainty $\sigma_{C}$ introduced by limited Monte Carlo statistics and binning effects in the determination of the detector efficiencies can be given as a function of $Q^2$
\begin{align}
\sigma_{C}(Q^{2})  & = \left( \sum_{i=0}^{7}\sum_{j=0}^{7}    \bar{Q}^{2i}     M^\text{cov}_{ij} \bar{Q}^{2j} \right)^{\frac{1}{2}}  = \left( \sum_{k=0}^{14} e_k  \bar{Q}^{2k} \right)^{\frac{1}{2}}
\end{align} 
which is obtained from the covariance matrix $M^\text{cov}_{ij} $ of the fit.
The coefficients $a_i$ and $e_i$ are given in Table~\ref{table:ModIndFunC}. 
\begingroup
\squeezetable
\begin{table}
\caption{Parameters describing the parameterization of the function $C$. The coefficients $a_i$ are obtained by fitting $C$ with an 7th-order polynomial. The coefficients $e_i$ are obtained from the error matrix of the fit. \label{table:ModIndFunC}}
\begin{ruledtabular}
\begin{tabular}{ c  c   c   c  c   c  c  c } 
\multicolumn{8}{ c }{Coefficients $C(Q^2)$ ($10^{-2}\,\mathrm{GeV}^{-2}c^4$)} \\ 
 $a_0$ & $a_1$ &  $a_2$ &  $a_3$ &  $a_4$ &  $a_5$ & $a_6$ & $a_7$ \\
 18.97 & -86.33 & 132.7 & -106.5 & 49.88 & -13.72 & 2.05 & -0.13 \\ \hline 
\multicolumn{8}{ c }{Coefficients  $\sigma^2_{C}(Q^2)$ ($10^{-5}\, \mathrm{GeV}^{-2}c^4$) } \\ 
$e_0$          & $e_1$         &  $e_2$          &  $e_3$                &  $e_4$     & $e_5$           & $e_6$       & $e_7$                     \\
1.933           &  -22.91       &119.3          & -363.2            & 725.6    &  -1011      & 1015    & -750.2               \\ 
$e_8$           & $e_9$         &$e_{10}$       &  $e_{11}$          &   $e_{12}$  &  $e_{13}$      & $e_{14}$                 \\
 410.8        & -166.3     &491.0       & -10.28              &   1.444        & -0.122            & 0.005                      \\ 
 \end{tabular}
\end{ruledtabular}
\end{table}
\endgroup